\documentclass[aps,prc,superscriptaddress,twoside,twocolumn,nofootinbib,showpacs]{revtex4-1}
\usepackage{amsmath,amssymb}
\usepackage{graphicx}
\usepackage{braket}

\begin{document}

\title{A model-independent analysis of $p + p \to pp(^1S_0) + \gamma$}

\author{K. Nakayama}
\email{nakayama@uga.edu}
\affiliation{Department of Physics and Astronomy, University of Georgia, Athens, GA 30602, USA }
\affiliation{\mbox{Institut f{\"u}r Kernphysik and J\"ulich Center for Hadron Physics, Forschungszentrum J{\"u}lich, 52425 J{\"u}lich, Germany}}

\author{F. Huang}
\email{huang@physast.uga.edu}
\affiliation{Department of Physics and Astronomy, University of Georgia, Athens, GA 30602, USA }

\begin{abstract}
The $pp$ hard bremsstrahlung reaction, $p + p \to pp(^1S_0) + \gamma$, in which
the two protons in the final state are in the $^1S_0$ state, is investigated
theoretically. Here, the most general spin structure of the $NN$ hard bremsstrahlung
reaction, consistent with symmetry principles, is derived from a partial-wave expansion
of this amplitude. Based on this spin structure, it is shown that there are only four
independent spin matrix elements in this reaction, which is a direct consequence of
reflection symmetry in the reaction plane. It is also shown that it requires at least eight
independent observables to determine them uniquely. The present method provides the
coefficients multiplying each spin operator in terms of the partial-wave or, equivalently,
multipole amplitudes. Some observables are expressed explicitly in terms of these
multipoles and a partial-wave analysis is performed.
The results should be useful in the analyses of the experimental data on the
$p + p \to pp(^1S_0) + \gamma$ reaction being taken, in particular, at the COSY
accelerator facility, as well as in providing some theoretical guidance to
the future experiments in this area.
\end{abstract}


\maketitle

\section{Introduction}
\label{sec:introduction}

The nucleon-nucleon ($NN$) bremsstrahlung and the related photo-absortion
reactions are among the most fundamental processes in hadron physics and, as
such, they have been used as tools to probe the hadron and/or quark dynamics
in a variety of strongly interacting systems. As an elementary process, the
photo-disintegration reaction $\gamma + d \to p + n$ has been studied widely and
for many years to probe the properties of the $NN$ interaction at short distances.
The $NN$ bremsstrahlung reaction, on the other hand, has been investigated
intensively in the past, mainly, to learn about the off-shell properties of the $NN$
interaction.  Unfortunately, however, in field theories off-shell properties are
unmeasurable quantities. Recently, in the absence of free bound diproton, hard  bremsstrahlung in $p + p \to pp(^1S_0) + \gamma$ has been measured for the first time by the COSY-ANKE Collaboration \cite{ANKE1,ANKE2}
as an alternative to the $\gamma + pp(^1S_0) \to p + p$  process; the latter is
complementary to the deuteron photo-disintegration process. Here, the hardness
of the bremsstrahlung is due to the fact that the invariant mass of the two protons
in the final state is constrained experimentally to be less than 3 MeV above
its minimum value of twice the proton mass. In this kinematic regime, the two
protons in the final state are practically confined to the $^1S_0$ state and most of
the available energy is carried by the bremsstrahlung. Therefore, this kinematic
regime is maximally away from the soft-photon limit. The proton-proton ($pp$)
hard bremsstrahlung reaction has been also measured at CELSIUS-Uppsala
\cite{JW09}. In spite of extensive studies of the $NN$ bremsstrahlung reaction in
the past, no dedicated experiments of $pp$ hard bremsstrahlung with the diproton
in the final state were available until the measurements of
Refs.~\cite{ANKE1,ANKE2,JW09}. Also, theoretical investigation of the $p + p \to
pp(^1S_0) + \gamma$ reaction is virtually inexistent so far.

In the present work, we derive the most general spin structure of the reaction
amplitude for the $NN$ bremsstrahlung process with the final two nucleons in the
$^1S_0$ state, $N + N \to NN(^1S_0) + \gamma$, following closely the method of
Ref.~\cite{NL05}.  We will show that there are four independent spin amplitudes
in this reaction and that it requires at least eight independent observables to
determine them. Usually, the structure of a transition amplitude is derived based
solely on symmetry principles. The present method also provides explicit formulae
for the coefficients multiplying each spin structure in terms of the partial-wave or,
equivalently, multipole amplitudes. This should be particularly useful in
model-independent analyses based on the partial-wave expansion of the reaction
amplitude.  Indeed, we shall show that some of the spin observables involving the
spins of the two nucleons in the initial state, together with the unpolarized cross
sections,  can determine the low angular momentum multipoles and/or their combinations.
These observables can be measured at the currently existent facilities such as
COSY-ANKE at J\"ulich.

The present paper is organized as follows: in Sec.~\ref{sec:spinstructure} the most
general spin structure for the $N + N \to NN(^1S_0) + \gamma$ reaction is derived. In
Sec.~\ref{sec:observables}, the coefficients multiplying each of the structures of the
bremsstrahlung amplitude, which are expressed in terms of the partial wave matrix
elements, are related to spin observables. In Sec.~\ref{sec:multipoles}, we establish
the relationship between the partial wave and multipole amplitudes. Sec.~\ref{sec:multJ2}
is dedicated to a partial-wave analysis of the $p + p \to pp(^1S_0) + \gamma$
reaction. Finally, Sec.~\ref{sec:summary} contains the summary. Some of the details of
the derivations are given in Appendices A and B.

\section{Spin structure of the $p + p \to pp(^1S_0) + \gamma$ amplitude}
\label{sec:spinstructure}

We start by making a partial-wave expansion of the $N + N \to N + N + \gamma$ reaction
amplitude. For completeness, we allow the photon to be virtual.  We have
\begin{align}
\bra{1m_s; S'M_{S'}}& \hat M(\vec k, \vec p\ '; \vec p)\ket{SM_S}  = \nonumber \\
% \equiv \bra{S'M_{S'}}\vec\epsilon_{\lambda'}\cdot\vec J\ket{SM_S}  \nonumber \\
&=\sum i^{L-L'-l} (SM_SLM_L|JM_J)  Y^*_{LM_L}(\hat p) \nonumber \\
&\ \ \ \ \ \ \times (S'M_{S'}L'M_{L'}|J'M_{J'})  Y_{L'M_{L'}}(\hat p') \nonumber\\
&\ \ \ \ \ \ \times (1m_s lm_l|jm_j)(jm_j J'M_{J'}|JM_J) \nonumber \\
&\ \ \ \ \ \ \times   Y_{lm_l}(\hat k) \tilde{M}^{J'S'JS}_{lL'L}(k, p'; p)   \ ,
\label{PWE_brems}
\end{align}
where
$S, L, J$ stand for the total spin, total orbital angular momentum and the total
angular momentum, respectively, of the initial $NN$ state. $M_S$, $M_L$ and
$M_J$ denote the corresponding projection quantum numbers. The primed
quantities stand for the quantum numbers of the final $NN$ state. $\ket{1m_s}$
stands for the spin state of the emitted photon; $l$ and $j$ denote the orbital
and total angular momentum of the emitted photon, respectively, relative to the
center-of-mass of the final $NN$ system. $m_l$ and $m_j$ denote the
corresponding projections, respectively. The summation runs over all the quantum
numbers not specified in the left-hand-side (l.h.s.) of Eq.(\ref{PWE_brems}).
$\vec p$ and $\vec p\ '$ denote the relative momenta of the two nucleons in the
initial and final states, respectively. $\vec k$ denotes the momentum of the emitted
photon with respect to the center-of-mass of the two nucleons in the final state.

Equation~(\ref{PWE_brems}) can be inverted to solve for the partial-wave matrix
element $\tilde{M}^{S'J'SJ}_{lL'L}(k, p'; p)$; we obtain
\begin{align}
\tilde{M}^{J'S'JS}_{lL'L}(k, p'; p) & =
\sum i^{L'+l-L} (SM_SL0|JM_J) \nonumber \\
&\times (S'M_{S'}L'M_{L'}|J'M_{J'}) \nonumber \\
&\times (1m_s lm_l|jm_j)(jm_jJ'M_{J'}|JM_J) \nonumber\\
& \times  \frac{8\pi^2}{2J+1} \sqrt{\frac{2L+1}{4\pi}}
\int d\Omega_{p'} Y_{L'M_{L'}}^*(\hat p') \nonumber \\
& \times
\int_{-1}^{+1} d(\cos\theta_k) Y_{lm_l}^*(\theta_k, 0)  \nonumber\\
&\times \braket{1m_s; S'M_{S'}|\hat M(\vec k, \vec p\ '; \vec p)|SM_S} \ ,
\label{pwme_brems}
\end{align}
where, without loss of generality, the $z$-axis is chosen along $\vec p$ and,
$\vec p$ and $\vec k$ define the $xz$-plane (the reaction plane);
$\cos\theta_k = \hat k \cdot \hat p$. The summation is over all the quantum
numbers not specified in the l.h.s. of the equality.

Following the method of Ref.~\cite{NL05}, the most general spin structure of the
transition operator can be extracted from Eq.(\ref{PWE_brems}) as
\begin{align}
\hat M(\vec k, \vec p\ '; \vec p) & = \sum \ket{1m_s; S'M_{S'}} \nonumber \\
& \ \ \ \ \times \braket{1m_s; S'M_{S'}|\hat M(\vec k, \vec p\ '; \vec p)|SM_S}\bra{SM_S} \ ,
\label{SSTRUC0_brems}
\end{align}
where the summation is over all the spin quantum numbers.

Inserting Eq.(\ref{PWE_brems}) into Eq.(\ref{SSTRUC0_brems}) and re-coupling gives
\begin{align}
\hat M(&\vec k, \vec p\ '; \vec p) = \sum i^{L-L'-l} (-)^{j+J'} [j J'][J]^2 \nonumber\\
&\times \tilde{M}^{J'S'JS}_{lL'L}(k, p'; p) \nonumber \\
&\times \sum_{\alpha\beta\gamma} (-)^\gamma [\gamma\beta]
  \begin{Bmatrix}
  S' & L' & J' \\   1 & l & j \\  \gamma & \beta & J
  \end{Bmatrix}
  \begin{Bmatrix}
  \gamma & \beta & J \\   L & S & \alpha
  \end{Bmatrix} \nonumber\\
&\times
[ B_S \otimes [A_{S'}\otimes \epsilon\ ]^\gamma ]^\alpha \cdot
[ Y_L(\hat p) \otimes [Y_{L'}(\hat p') \otimes Y_{l}(\hat k)]^\beta ]^\alpha  \ ,
\label{SSTRUC1_brems}
\end{align}
where we have introduced the notations
$\epsilon_{1m_s} \equiv \ket{1m_s}$, $A_{S'M_{S'}} \equiv \ket{S'M_{S'}}$, and
$B_{SM_S} \equiv (-)^{S-M_S}\bra{S-M_{S}}$, in addition to
\begin{align}
[j] & \equiv \sqrt{2j+1} \ , \nonumber \\
[j_1 ... j_n] & \equiv [j_1]...[j_n] \ .
\label{notation1}
\end{align}

In the following we restrict ourselves to $NN$ bremsstrahlung with the final $NN$ subsystem in the
$^1S_0$ state ($S'=L'=J'=0$). Eq.(\ref{SSTRUC1_brems}) then reduces to
\begin{align}
\hat M(\vec k, \vec p\ '; \vec p) &= \frac{1}{\sqrt{4\pi}} \sum i^{L-l} (-)^{-J+1} [J]^2
M^{JS}_{lL}(k, p'; p)  \nonumber\\
&\times \sum_{\alpha}
  \begin{Bmatrix}
  S & L & J \\   l & 1 & \alpha
  \end{Bmatrix} \nonumber \\
&\times
[B_S \otimes [A_{S'=0}\otimes \epsilon]^1 ]^\alpha \cdot
[Y_L(\hat p) \otimes Y_{l}(\hat k)]^\alpha \ ,
\label{SSTRUC2_brems}
\end{align}
where $\tilde{M}^{JS}_{lL}\equiv \tilde{M}^{00JS}_{l0L}$.

We now expand $[ B_S \otimes [A_0\otimes\epsilon]^1 ]^\alpha$, for each tensor rank $\alpha$,
in terms of the complete set of available spin operators in the problem, i.e., the polarization
of the photon $\vec\epsilon$, and the Pauli spin matrices $\vec\sigma_1$ and $\vec\sigma_2$
corresponding to the two nucleons together with the corresponding spin identity matrices.
Note that the final $NN$ subsystem is in the $^1S_0$ state ($S'=0$).
There are, then, four cases to be considered:
\begin{align}
S=0, \alpha=1 & : \
[B_S \otimes [A_0\otimes\epsilon]^1]^1  =  \vec\epsilon P_{S=0} \ , \nonumber \\
S=1, \alpha=0 & : \
[B_S \otimes [A_0\otimes\epsilon]^1]^0  = \frac{1}{\sqrt{3}} \vec\epsilon\cdot \vec T_{S=1} \ ,
\nonumber \\
S=1, \alpha=1 & : \
[B_S \otimes [A_0\otimes\epsilon]^1]^1  =  \frac{i}{\sqrt{2}}
\left(\vec\epsilon\times \vec T_{S=1}\right) \ , \nonumber \\
S=1, \alpha=2 & : \
[B_S \otimes [A_0\otimes\epsilon]^1]^2  = - [\vec\epsilon \otimes \vec T_{S=1}]^2 \ ,
\label{SOPER}
\end{align}
where $P_{S}$ stands for the spin projection operator onto the (initial) spin
singlet and triplet state, respectively, as $S=0$ or 1. $\vec T_S$ is the
spin-flip operator from the initial $NN$ spin state $S$ to the final $NN$ spin
state $S'=S\pm 1$. In terms of the Pauli spin matrices, they are given by
\begin{align}
P_S & \equiv \frac{1}{4}\left[2S + 1 - (-)^S \vec\sigma_1\cdot\vec\sigma_2\right] \ , \nonumber \\
\vec T_{S} &\equiv \frac{1}{2}\left(\vec\sigma_1 - \vec\sigma_2\right)P_S
= \frac{1}{4}\left[\left(\vec\sigma_1 - \vec\sigma_2\right)
 + (-)^S i \left(\vec\sigma_1 \times \vec\sigma_2\right)\right] \ .
\label{PROJ}
\end{align}
In Eq.(\ref{SOPER}), the numerical factors on the right-hand-side (r.h.s.) of the
equalities are such that the spin matrix elements of the l.h.s equal the
corresponding ones of the r.h.s. For this purpose, we have made use of the
formulas
\begin{align}
\left[\vec A \otimes \vec B \right]^1 & = \frac{i}{\sqrt{2}} \left(\vec A \times \vec B \right) \ , \nonumber \\
\vec A \cdot \vec T_{S=1}\ket{SM_S} & = \delta_{S,1} A_{M_S} \ket{00} \ .
\label{f1}
\end{align}
for arbitrary vectors $\vec A$ and $\vec B$.

With the quantization axis $\hat z$ chosen along $\hat p$,
$[Y_{L}(\hat p) \otimes Y_l(\hat k)]^\alpha$ in Eq.(\ref{SSTRUC2_brems})
can be expressed as
\begin{align}
[Y_{L}(\hat p) \otimes Y_l(\hat k)]^0 & =  0 \ , \nonumber\\
[Y_{L}(\hat p) \otimes Y_l(\hat k)]^1 & =  \frac{[Ll]}{4\pi} \left(1-\delta_{L0}\right)
\Big[ \sqrt{\frac{2}{l(l+1)}}(L0l1|11)\nonumber\\[1ex]
\times P^1_{l}(\hat k \cdot \hat p) \hat n_1 + &
(L0l0|10) P_{l}(\hat k \cdot \hat p) \hat p \Big]
 + \frac{\sqrt{3}}{4\pi}\delta_{L0}\delta_{l1} \hat k\ , \nonumber\\
[Y_{L}(\hat p) \otimes Y_l(\hat k)]^2 & =
 a_{lL} [\hat p \otimes \hat n_2]^2 + b_{lL} [\hat k \otimes \hat n_2]^2 \ ,
\label{AOPER}
\end{align}
where
\begin{align}
\hat n_1 & \equiv [(\hat p \times \hat k) \times \hat p]/|\hat p \times \hat k| \ , \nonumber \\
\hat n_2 &\equiv (\hat p \times \hat k)/|\hat p \times \hat k| \ .
\label{n1n2}
\end{align}
Note that total parity conservation demands that $(-)^{L+l}=-1$. Also, in the
second equality in Eq.~(\ref{AOPER}), we have explicitly isolated the $L=0$
state contribution for further convenience. The coefficients $a_{lL}$ and $b_{lL}$
are calculated explicitly in Appendix A.

Inserting Eqs.(\ref{SOPER},\ref{AOPER}) into Eq.(\ref{SSTRUC2_brems}) and
using the identity
\begin{equation}
3 [\ \vec\epsilon \otimes \vec T_{S}\ ]^2 \cdot [\ \hat a \otimes
\hat b\ ]^2 = \frac{3}{2} [\ \vec\epsilon \cdot \hat a \vec T_{S} \cdot \hat b
+ \vec\epsilon \cdot \hat b \vec T_{S} \cdot \hat a \ ]
- (\hat a \cdot \hat b) \vec\epsilon \cdot \vec T_{S} \ ,
\label{tensor}
\end{equation}
with $\hat a$ and $\hat b$ denoting arbitrary unit vectors, we have
\begin{align}
& \hat M(\vec k, \vec p\ '; \vec p) =  \left[F_1 \vec\epsilon\cdot\hat k +
F_2 \vec\epsilon\cdot\hat p
+  F_3 \vec\epsilon\cdot\hat n_1 \right] P_{S=0} \nonumber \\
&+ \left[ iF_4 (\vec\epsilon\times\hat k)
+ iF_5 \left(\vec\epsilon\times\hat p\right)
+ iF_6 \left(\vec\epsilon\times\hat n_1 \right) \right. \nonumber\\
& \left. + F_7 \left(\vec\epsilon\cdot\hat p \hat n_2 +
\vec\epsilon\cdot\hat n_2 \hat p\right)
+ F_8 (\vec\epsilon\cdot\hat k \hat n_2 +
\vec\epsilon\cdot\hat n_2 \hat k)\right]\cdot\vec T_{S=1} \ ,
\label{SSTRUC4_brems}
\end{align}
where the coefficients $F_i\ (i=1,...,8)$ are given in Eq.~(\ref{COEFF_brems}) below.
The above result is the most general spin structure of the $NN$ bremsstrahlung
amplitude consistent with symmetry principles when the final $NN$ subsystem is in
the $^1S_0$ state. The first three terms are the central $NN$ spin singlet transitions.
The fourth, fifth, and sixth terms are tensors of rank 1 describing the $NN$ spin
triplet$\to$singlet transitions. The last two terms involving the coefficients $F_7$ and
$F_8$ correspond to the tensor interaction of rank 2. Apart from the first three terms,
all the other terms are spin-flip transitions.

Here we note that the gauge freedom always allows one to make the choice for the photon
polarization $\epsilon_\mu = \epsilon'_\mu - \eta k_\mu$ for any value of the gauge
parameter $\eta$. For example, with the choice $\eta=\epsilon'_0/k_0$, which leads
to $\epsilon_0=0$, the photon polarization has no scalar component and the reaction
amplitude contains a virtual photon with transverse and longitudinal components.
Eq.~(\ref{SSTRUC4_brems}) corresponds to this choice. Alternatively, one
could make the choice $\eta = (\vec\epsilon\ '\cdot\hat k)/|\vec k|$ for the gauge
parameter, leading to $\vec\epsilon\cdot\hat k = 0$ but $\epsilon_0 \ne 0$, i.e., the
virtual photon has no longitudinal component and the reaction amplitude would
correspond to a virtual photon with transverse and scalar components. Of course,
amplitudes with different gauge choices are equivalent to each other by gauge
invariance.

The coefficients multiplying the spin operators in Eq.~(\ref{SSTRUC4_brems})
contain the dynamics of the reaction process and are given by
\begin{align}
F_1 & =  \frac{1}{(4\pi)^{\frac{3}{2}}} i\delta_{L0}\delta_{l1}\delta_{J0}
\tilde{M}^{00}_{10}(k, p'; p) \ , \nonumber \\
F_2 & =  \frac{1}{(4\pi)^{\frac{3}{2}}}\frac{1}{\sqrt{3}}
\sum i^{L-l}(-)^{l}[L]^2 [l](L0l0|10) \left(1-\delta_{L0}\right) \nonumber \\
&\ \ \ \ \ \ \ \ \ \ \ \ \ \ \ \ \ \ \ \
\times \tilde{M}^{L0}_{lL}(k, p'; p)P_l(\hat k \cdot \hat p) \ , \nonumber \\
F_3 & =  \frac{1}{(4\pi)^{\frac{3}{2}}}\frac{1}{\sqrt{3}}
 \sum i^{L-l}(-)^{l}[L]^2 [l]
\sqrt{\frac{2}{l(l+1)}}(L0l1|11) \nonumber \\
&\ \ \ \ \ \ \ \ \ \ \ \ \ \ \ \ \ \ \ \
\times \left(1-\delta_{L0}\right) \tilde{M}^{L0}_{lL}(k, p'; p)P^1_l(\hat k \cdot \hat p) \ , \nonumber \\
F_4 & = \frac{1}{(4\pi)^{\frac{3}{2}}} \sqrt{\frac{3}{2}}
\delta_{L0}\delta_{l1}\delta_{J1} \tilde{M}^{11}_{10}(k, p'; p) \ , \nonumber \\
F_5 & =  \frac{1}{(4\pi)^{\frac{3}{2}}} \frac{1}{\sqrt{2}}
\sum  i^{L-l}(-)^{-J}[J]^2 [Ll](L0l0|10)\nonumber \\
&\ \ \ \ \ \ \ \times \left(1-\delta_{L0}\right)
\begin{Bmatrix}
  1 & L & J \\   l & 1 & 1
  \end{Bmatrix}
\tilde{M}^{J1}_{lL}(k, p'; p)P_l(\hat k \cdot \hat p) \ , \nonumber \\
F_6 & =  \frac{1}{(4\pi)^{\frac{3}{2}}} \frac{1}{\sqrt{2}}
\sum  i^{L-l}(-)^{-J}[J]^2 [Ll] \sqrt{\frac{2}{l(l+1)}} \nonumber \\[1ex]
& \times (L0l1|11)  \left(1-\delta_{L0}\right)
 \begin{Bmatrix}
  1 & L & J \\   l & 1 & 1
  \end{Bmatrix}
\nonumber \\[1ex]
& \times \tilde{M}^{J1}_{lL}(k, p'; p)P^1_l(\hat k \cdot \hat p) \ , \nonumber \\
%\nonumber
%\end{align}
%\begin{align}
F_7 & =  \frac{1}{2\sqrt{4\pi}} \sum i^{L-l}(-)^{-J}[J]^2
  \begin{Bmatrix}
  1 & L & J \\   l & 1 & 2
  \end{Bmatrix} \nonumber\\
&\ \ \ \ \ \ \ \ \ \ \ \ \ \
\times \tilde{M}^{J1}_{lL}(k, p'; p) a_{lL} \ , \nonumber \\
F_8 & =  \frac{1}{2\sqrt{4\pi}} \sum i^{L-l}(-)^{-J}[J]^2
  \begin{Bmatrix}
  1 & L & J \\   l & 1 & 2
  \end{Bmatrix} \nonumber\\
&\ \ \ \ \ \ \ \ \ \ \ \ \ \
\times \tilde{M}^{J1}_{lL}(k, p'; p) b_{lL} \ .
\label{COEFF_brems}
\end{align}

The summations are over all the quantum numbers appearing explicitly in the r.h.s. of
the equalities. Note that they are restricted by total parity conservation, $(-)^{L+l}=-1$,
and the Pauli principle. In the case of $pp$ bremsstrahlung, the latter leads to
$(-)^{L+S} = 1$. These imply that the $pp$ spin triplet $\to$ singlet transitions involve
only the even photon orbital angular momenta, while the singlet $\to$ singlet transitions
involve only the odd orbital angular momenta.

Note that, in Eq.~(\ref{SSTRUC4_brems}), the terms proportional to $\vec\epsilon\cdot\vec k$
contribute only for virtual photons, such as in dilepton productions. Also, the coefficient $F_4$
is absent in $pp$ bremsstrahlung due to the Pauli principle.

Although the primary focus of the present work is on the hard bremsstrahlung production in
$pp$ collisions, all the results derived in this and the following two sections apply to $pn$ collisions
as well.

\section{Observables}
\label{sec:observables}

In this section, we will relate the coefficients $F_i$ in Eq.~(\ref{SSTRUC4_brems}) to the
corresponding spin matrix elements and then relate these matrix elements to some observables.
In what follows, we consider the real photon and take the two independent (linear) photon
polarizations to be parallel and perpendicular to the reaction plane,
\footnote{Note that for a circularly polarized photon, $\vec\epsilon_{\lambda'} \to \vec\epsilon\,^*_{\lambda'}$,
for the photon is being emitted in this reaction instead of being absorbed.}
\begin{align}
\vec\epsilon_\parallel &\equiv \cos\theta_k \hat x  - \sin\theta_k \hat z \ , \nonumber \\
\vec\epsilon_\perp &\equiv \hat y \ .
\label{photon-pol}
\end{align}

Then, using the short-hand notation
\begin{equation}
M^{\lambda'}_{SM_S} \equiv \braket{1\lambda'; 00|\hat M(\vec k, \vec p\ '; \vec p) |SM_S} \ ,
\label{shn}
\end{equation}
the spin matrix elements of the amplitude given by Eq.(\ref{SSTRUC4_brems}),
with a specific photon polarization state $\vec\epsilon_{\lambda'}$, are given by
\begin{align}
M^\parallel_{00}& = -F_2\sin\theta_k + F_3\cos\theta_k \ , \nonumber \\
M^\perp_{00}& = 0 \ ,
\label{spin-singlet}
\end{align}
for the transition of $NN$ spin singlet $\to$ singlet state, and
\begin{align}
M^\parallel_{10}& = 0 \ , \nonumber \\
M^\perp_{10}& = -i(F_4\sin\theta_k+F_6) + F_7 +F_8\cos\theta_k \ , \nonumber \\
M^\parallel_{1\pm 1}& = \frac{-1}{\sqrt{2}}
\left[F_4 + F_5\cos\theta_k + (F_6-iF_7)\sin\theta_k\right] \ , \nonumber \\
M^\perp_{1\pm 1}& = \mp \frac{1}{\sqrt{2}}\left[i(F_4\cos\theta_k+F_5) + F_8\sin\theta_k\right] \ ,
\label{spin-triplet}
\end{align}
for the triplet $\to$ singlet transition. Here, we made use of Eq.~(\ref{f1}).
It should be emphasized that Eqs.(\ref{spin-singlet},\ref{spin-triplet}) reveal
that there are only four independent spin matrix elements in this reaction. This is
a direct consequence of reflection symmetry in the reaction plane
\cite{Bohr,Satchler,NL04}. The results obtained in the remainder of this section and
in Sec.~\ref{sec:multJ2}, are consequences of this property of the spin matrix
elements.

In this work, we consider the unpolarized cross section, $d\sigma/d\Omega$, and
the spin observables involving the spins of the initial state nucleons only, namely,
the analyzing powers, $A_i$, and the spin correlation coefficients, $A_{ij}$. They are
defined as
\begin{align}
\frac{d\sigma}{d\Omega} & \equiv  \frac{1}{4} Tr[\hat M \hat M^\dagger] \ , \nonumber \\
\frac{d\sigma}{d\Omega} A_i & \equiv  \frac{1}{4} Tr[\hat M \sigma_i(1) \hat M^\dagger] \ , \nonumber \\
\frac{d\sigma}{d\Omega} A_{ij} & \equiv  \frac{1}{4} Tr[\hat M \sigma_i(1) \sigma_j(2) \hat M^\dagger] \ .
\label{obs}
\end{align}

Following Appendix D of Ref.~\cite{NL05}, the above observables are readily expressed
in terms of the spin matrix elements $M^{\lambda'}_{SM_S}$.  We have for unpolarized
cross section and (diagonal) spin correlation coefficients
\begin{align}
\frac{d\sigma}{d\Omega} &= \frac{1}{4}
\left(|M^\parallel_{00}|^2 + 2|M^\parallel_{11}|^2 +
      |M^\perp_{10}|^2 + 2|M^\perp_{11}|^2\right) \ , \nonumber \\
\frac{d\sigma}{d\Omega}A_{xx} &= \frac{1}{4}
\left(-|M^\parallel_{00}|^2 + 2|M^\parallel_{11}|^2 +
      |M^\perp_{10}|^2 - 2|M^\perp_{11}|^2\right) \ , \nonumber \\
\frac{d\sigma}{d\Omega}A_{yy} &= \frac{1}{4}
\left(-|M^\parallel_{00}|^2 - 2|M^\parallel_{11}|^2 +
      |M^\perp_{10}|^2 + 2|M^\perp_{11}|^2\right) \ , \nonumber \\
\frac{d\sigma}{d\Omega}A_{zz} &= \frac{1}{4}
\left(-|M^\parallel_{00}|^2 + 2|M^\parallel_{11}|^2 -
      |M^\perp_{10}|^2 + 2|M^\perp_{11}|^2\right) \ , \nonumber \\
\label{XSC-SCOR-unpol}
\end{align}
which can be inverted to yield
\begin{align}
|M^\parallel_{00}|^2 &= \frac{d\sigma}{d\Omega}
\left(1 - A_{xx} - A_{yy} - A_{zz}\right) \ , \nonumber \\
|M^\perp_{10}|^2 &= \frac{d\sigma}{d\Omega}
\left(1 + A_{xx} + A_{yy} - A_{zz}\right) \ , \nonumber \\
2|M^\parallel_{11}|^2 &= \frac{d\sigma}{d\Omega}
\left(1 + A_{xx} - A_{yy} + A_{zz}\right) \ , \nonumber \\
2|M^\perp_{11}|^2 &= \frac{d\sigma}{d\Omega}
\left(1 - A_{xx} + A_{yy} + A_{zz}\right) \ .
 \label{SPINMTX}
\end{align}
Therefore, the measurement of the cross section and the spin correlation
coefficients can determine the magnitude of the individual
spin matrix elements.

The off-diagonal spin correlation coefficients vanish identically as a
consequence of reflection symmetry, except for $A_{xz} (=A_{zx})$
which is given by
\begin{align}
\frac{d\sigma}{d\Omega} A_{xz} & = \frac{d\sigma}{d\Omega} A_{zx} \nonumber \\
& = - \frac{1}{\sqrt{2}}Re[M^\parallel_{00}{M^\parallel_{11}}^*
- M^\perp_{10}M^{\perp *}_{11}] \ .
\label{Axz}
\end{align}

The analyzing power reads
\begin{equation}
\frac{d\sigma}{d\Omega} A_y = \frac{1}{\sqrt{2}}Im[M^\parallel_{00}{M^\parallel_{11}}^*
- M^\perp_{10}M^{\perp *}_{11}]  \ .
\label{Ay}
\end{equation}
$A_x = A_z = 0$ by symmetry.

Equations~(\ref{Axz},\ref{Ay}) show that the analyzing power $A_y$, together with
the spin correlation coefficient $A_{xz}$, determines the relative phase of the
combination of the spin matrix elements involving the two independent photon
polarization states, $\vec\epsilon_\perp$ and $\vec\epsilon_\parallel$.
Note that both the real and imaginary parts are required to determine the relative
phase uniquely. It is also immediate from Eqs.~(\ref{Axz},\ref{Ay}) that one
obvious possibility to disentangle the combination of those two terms with different
photon polarization is to measure $A_{xz}$ and $A_y$ together with the
polarization of the emitted photon. This means that, in total, we need at least eight
independent observables to determine these spin amplitudes uniquely, apart from the
irrelevant overall phase.\footnote{A general discussion on the closely related issue of ``complete experiment" in pseudoscalar meson photoproduction has been given by Chiang and Tabakin \cite{CT95}.}
Evidently, a complete determination of the hard bremsstrahlung
amplitude poses an enormous experimental challenge.

It is also interesting to note another consequence of reflection symmetry that
relates the double polarization observable with the single polarization observable;
namely, \cite{NL04}
\begin{equation}
A_{yy} =  - \Sigma \ ,
\label{eq:AyySigma}
\end{equation}
where $\Sigma$ stands for the outgoing photon asymmetry given by
\begin{equation}
\frac{d\sigma}{d\Omega}\Sigma \equiv
\frac{d\sigma^\parallel}{d\Omega} -
 \frac{d\sigma^\perp}{d\Omega} \ ,
\label{Photon-Asym}
\end{equation}
with $\sigma^{\lambda'}$ denoting the cross section with a given photon
polarization $\vec\epsilon_{\lambda'}$. Eq.~(\ref{eq:AyySigma}) can also
be verified simply by an inspection of the result for $A_{yy}$ in
Eq.~(\ref{XSC-SCOR-unpol}) together with the definition of $\Sigma$ given
in the above equation.

Some combinations of the spin matrix elements can be also extracted
from the measurements of the transverse spin correlation coefficients
$A_{xx}$ and $A_{yy}$ in conjunction with the unpolarized cross section.
The coefficient $A_{zz}$, which involves the longitudinal polarization of
the beam and target nucleon spins, are experimentally more challenging
to be measured than the transverse coefficients. We have, from
Eq.~(\ref{XSC-SCOR-unpol})
\begin{align}
2\frac{d\sigma}{d\Omega} \left( A_{xx} + A_{yy} \right) & = |M^\perp_{10}|^2 - |M^\parallel_{00}|^2 \ , \nonumber \\
2\frac{d\sigma}{d\Omega} \left( A_{xx} - A_{yy} \right) & = 2|M^\parallel_{11}|^2 - 2|M^\perp_{11}|^2 \ , \nonumber \\
2\frac{d\sigma}{d\Omega} \left( 1 + A_{xx} \right) & = 2|M^\parallel_{11}|^2 + |M^\perp_{10}|^2 \ , \nonumber \\
2\frac{d\sigma}{d\Omega} \left( 1 + A_{yy} \right) & = 2|M^\perp_{11}|^2 + |M^\perp_{10}|^2 \ .
\label{Atrans}
\end{align}
\section{Multipole amplitudes.}
\label{sec:multipoles}

The bremsstrahlung amplitude can also be decomposed in terms of more familiar
multipole amplitudes instead of the partial wave expansion given by
Eq.(\ref{COEFF_brems}). The multipole decomposition of the inverse process,
$\gamma + pp(^1S_0) \to p + p$, has been given elsewhere \cite{WNA96}.
In this section we will establish a
relationship between the multipole and partial-wave amplitudes introduced in the
previous section. We do this in the overall c.m. frame of the system with the coordinate
system as chosen in Sec.~\ref{sec:spinstructure} (see below Eq.~(\ref{pwme_brems})).

The matrix element for the transition $p + p \to pp(^1S_0) + \gamma$ in the
helicity basis  is then related to that of the plane-wave basis by \cite{JW59}
\begin{align}
& \braket{1\lambda'|\hat M(\vec k, \vec p\,'; \vec p)|S \lambda} = \nonumber \\
& \ \ \ \ \ \ \ \ \ \ \ \ \ \ \ \ \ = \sum_{m_s} D^{1\,^*}_{m_s\lambda'}(R)
\braket{ 1m_s|\hat M(\vec k, \vec p\ '; \vec p)| S M_S} \ ,
\label{eq:g15}
\end{align}
where $D^1_{m_s\lambda'}(R)$ stands for the spin rotation matrix
which rotates our coordinate frame by an angle $\theta_k$ about the y-axis
$(\alpha = \gamma =0, \beta=\theta_k)$ such that the new $z'$-axis is
along the photon momentum $\vec k$. Note that, since $\vec p$ is along the
positive $z$-axis, the initial plane-wave state coincides with the helicity state,
i.e., $\lambda = M_S$. In the above equation and in what follows, we have
suppressed the reference to the spin ($S'=M_{S'}=0$) of the two nucleons in the
final state.

Now, the partial wave decomposition of the plane-wave matrix element given by
Eq.~(\ref{PWE_brems}) for the case of the final two nucleons in the $^1S_0$
state reduces to
\begin{align}
&\braket{ 1 m_s|\hat M(\vec k, \vec p\ '; \vec p)| SM_S} =
\frac{1}{4\pi} \sum [L] (lm_l1m_s|jM_S) \nonumber \\
& \ \ \ \ \ \ \ \ \ \ \ \ \ \ \ \ \ \ \ \ \ \times (L0SM_S|jM_S) Y_{lm_l}(\hat k) M^{jS}_{lL}(k, p'; p) \ ,
\label{eq:g2}
\end{align}
where the summation runs over all the quantum numbers not appearing on the l.h.s of the
equality. In the above equation, we have redefined the partial wave
matrix elements to incorporate a phase, i.e.,
\begin{equation}
M^{jS}_{lL}(k, p'; p) \equiv i^{L-l}(-)^S \tilde{M}^{jS}_{lL}(k, p'; p)
\label{eq:g2a}
\end{equation}
for further convenience. Note also the change in the angular momentum coupling scheme
in Eq.~(\ref{eq:g2}) compared to Eq.~(\ref{PWE_brems}).  The reason for this change is to
have a close comparison of the results of this section with those in Ref.~\cite{WNA96}.

Inserting Eq.~(\ref{eq:g2}) into Eq.~(\ref{eq:g15}), and using the relations \cite{JW59}
\begin{align}
Y_{l\mu}(\hat k) & = \frac{[l]}{\sqrt{4\pi}} D^{l\,^*}_{\mu 0}(R) \ , \nonumber \\
(l01\lambda'|j\lambda') D^{j\,^*}_{\lambda \lambda'}(R) & =
\sum_{\mu\mu'} (l\mu1\mu'|j\lambda)
D^{l\,^*}_{\mu 0}(R) D^{1\,^*}_{\mu' \lambda'}(R) \ ,
\label{eq:g4}
\end{align}
yields
\begin{align}
& \braket{ 1\lambda'|\hat M(\vec k, \vec p\ '; \vec p) |S\lambda} =
 \sum_{jL} \frac{[L]}{(4\pi)^{\frac{3}{2}}}(L0S\lambda|j\lambda)d^j_{\lambda\lambda'}(\theta_k) \nonumber \\[1ex]
& \ \ \ \ \ \ \ \ \ \ \ \ \ \ \ \ \ \ \ \ \ \ \ \ \ \ \ \ \ \times \sum_{l}[l] (l01\lambda'|j\lambda')
M^{jS}_{lL}(k, p'; p) \ .
\label{eq:g5}
\end{align}
Note that, in our case, $D^{j\,^*}_{\lambda\lambda'}(R) = d^{j\,^*}_{\lambda\lambda'}(\theta_k)=
d^{j}_{\lambda\lambda'}(\theta_k)$.

Now, the electric ($E$) and magnetic ($M$) multipoles are classified
according to $(-)^{j+l+1} =+ 1$ and $(-)^{j+l+1}=-1$, respectively.
In addition, they are transverse amplitudes, i.e., for $\lambda' =\pm 1$
only. The longitudinal multipole amplitude ($L$) corresponds to $\lambda'=0$.
Then, due to the properties of the Clebsh-Gordon
coefficient $(l01\lambda'|j\lambda')$,
\begin{align}
(l01\lambda'|j\lambda') & = (-)^{l+1+j-2\lambda'}(l01-\lambda'|j-\lambda') \ , \nonumber \\
(l010|j0) & = 0  \ \  \rm{if} \ \  \it{l}+\rm{1}+\it{j}=\rm{odd} \ ,
\label{eq:g6}
\end{align}
one sees that Eq.~(\ref{eq:g5}) can be rewritten as
\begin{align}
\bra{1\lambda'} &\hat M(\vec k, \vec p\ '; \vec p)\ket{S\lambda} = \nonumber \\
& = \sum_{jL} \frac{[L]}{(4\pi)^{\frac{3}{2}}} (L0S\lambda|j\lambda)
d^{j}_{\lambda\lambda'}(\theta_k)
\left[ \delta_{\lambda' 0}L_{jSL}(k, p'; p) \right. \nonumber \\
&\ \ \ \ \ \  \ \ \left. + |\lambda'|E_{jSL}(k, p'; p) + \lambda' M_{jSL}(k, p'; p)\right] \ ,
\label{eq:g7}
\end{align}
with
\begin{align}
L_{jSL} & \equiv \sum_{l} [l] (l010|j0) M^{jS}_{lL} \ , \nonumber \\
E_{jSL} & \equiv \sum_{l} \left( \frac{1+(-)^{l+1+j}}{2}\right)[l]
(l011|j1) M^{jS}_{lL} \ , \nonumber \\
M_{jSL} & \equiv \sum_{l} \left(\frac{1-(-)^{l+1+j}}{2}\right)[l]
(l011|j1) M^{jS}_{lL} \ ,
\label{eq:g8}
\end{align}
where the momentum arguments of the amplitudes have been omitted.
The above equations relate the multipole amplitudes to the partial wave
amplitudes.

For a real (transverse) photon no longitudinal multipoles exist. Therefore,
Eq.~(\ref{eq:g7}) reduces to
\begin{align}
&\braket{ 1\lambda'|\hat M(\vec k, \vec p\ '; \vec p)| S\lambda} =
\sum_{jL} \frac{[L]}{(4\pi)^{\frac{3}{2}}} (L0S\lambda|j\lambda)
d^{j}_{\lambda\lambda'}(\theta_k) \nonumber \\[1ex]
&\ \ \ \ \ \ \ \ \ \ \ \ \ \ \ \ \ \ \ \ \ \ \ \
\times \left[ E_{jSL}(k, p'; p) + \lambda' M_{jSL}(k, p'; p)\right] \ ,
\label{eq:g9}
\end{align}
which agrees with Eq.~(2) of Ref.~\cite{WNA96} apart from an overall normalization
factor.

Equation~(\ref{eq:g7}) can be readily inverted to solve for the multipole amplitudes. We
have
\begin{align}
L_{jSL} & = [L] \sum_\lambda (L0S\lambda|j\lambda) A^{jS}_{0,\lambda} \ , \nonumber \\
E_{jSL} & = [L] \sum_\lambda (L0S\lambda|j\lambda)
                    \frac{1}{2} \left[ A^{jS}_{+1,\lambda} + A^{jS}_{-1,\lambda}\right] \ , \nonumber \\
M_{jSL} & = [L] \sum_\lambda (L0S\lambda|j\lambda)
                    \frac{1}{2} \left[ A^{jS}_{+1,\lambda} - A^{jS}_{-1,\lambda}\right] \ ,
\label{eq:g10}
\end{align}
where
\begin{equation}
A^{jS}_{\lambda',\lambda} \equiv \sqrt{4\pi} \int d\Omega_k\  d^j_{\lambda\lambda'}(\theta_k)
\braket{ 1\lambda'|\hat M(\vec k, \vec p\ '; \vec p)| S\lambda} \ .
\label{eq:g11}
\end{equation}

Eq.~(\ref{eq:g15}) can be inverted to obtain
\begin{align}
\bra{ 1m_s} & \hat M(\vec k, \vec p\ '; \vec p)\ket{S M_S} = \nonumber \\
& = \sum_{\lambda'} D^1_{\lambda' m_s}(R)
\braket{ 1\lambda'|\hat M(\vec k, \vec p\ '; \vec p)| S \lambda} \ .
\label{eq:g16}
\end{align}
Inserting the multipole expansion of the helicity amplitude, Eq.~(\ref{eq:g7}),
into the above equation yields
\begin{align}
\bra{1m_s} \hat M(\vec k, \vec p\ '&; \vec p)\ket{S M_S} =
\frac{1}{(4\pi)}\sum [L] (lm_l1m_s|jM_S) \nonumber \\
& \times (L0SM_S|jM_S) Y_{lm_l}(\hat k)
\frac{[l]}{[j]^2} (l01\lambda'|j\lambda')  \nonumber \\
& \times \left[ \delta_{\lambda' 0}L_{jSL}(k, p'; p) \right. \nonumber \\
& \ \ \ \left. + |\lambda'|E_{jSL}(k, p'; p) + \lambda' M_{jSL}(k, p'; p)\right] ,
\label{eq:g17}
\end{align}
where the summation runs over all the quantum numbers not appearing in the l.h.s.
of the equality. Recall that $\lambda'=M_S$. In arriving at the above result, we have
made used of the property \cite{JW59}
\begin{align}
d^{j'}_{\lambda\lambda'}d^{j}_{\mu\mu'} &= \sum_{k m m'}
(j'\lambda j-\mu|km)(j'\lambda' j-\mu'|km') \nonumber \\[1ex]
& \ \ \ \ \ \ \ \ \times (-)^{\mu-\mu'}d^{k}_{m,m'} \nonumber \\
& = \sum_{k m m'} (-)^{2j+2k+\lambda + \lambda'} \frac{[k]^2}{[j]^2} \nonumber \\
& \ \ \ \ \ \ \times (k-m j'\lambda|j\mu)(k-m' j'\lambda'|j\mu') d^{k}_{m,m'} \ .
\label{eq:f18}
\end{align}

Equation~(\ref{eq:g17}) gives a direct decomposition of the hard bremsstrahlung amplitude
in the plane-wave basis in terms of the multipole amplitudes.

Finally, comparing Eqs.~(\ref{eq:g2},\ref{eq:g17}), we have
\begin{align}
M^{jS}_{lL} = \frac{[l]}{[j]^2} \sum_{\lambda'}& (l01\lambda'|j\lambda')  \nonumber \\
& \times \left[ \delta_{\lambda' 0}L_{jSL} + |\lambda'|E_{jSL} + \lambda' M_{jSL}\right] ,
\label{eq:g18}
\end{align}
which gives the partial wave amplitudes in terms of the multipole amplitudes.

\section{Partial-wave analysis of the $p + p \to pp(^1S_0) + \gamma$ reaction}
\label{sec:multJ2}

The $\gamma + pp(^1S_0) \to p + p$ reaction has been analyzed in terms of few multipoles
with $j\le 2$ in the past \cite{WNA96}. Here, we perform a similar analysis of the $p + p \to
pp(^1S_0) + \gamma$ reaction, by extending it to include the spin observables which have
not been considered in that reference.  In addition, we will keep the $^3F_2 \to{^1S_0d}$
partial wave state in the analysis. This state, which contributes to the $M_{213}$ multipole,
has been ignored in Ref.~\cite{WNA96}.

Restricting the partial-wave expansion in Eq.(\ref{COEFF_brems}) to $J (=j) \le 2$,
we have the following partial wave states:
\footnote{Here we adopt the same notation for the partial-wave states as used in Ref.~\cite{Meyer},
i.e., $^{2S+1}L_J \rightarrow \ ^{2S'+1}L'_{J'} l$, where  $S$, $L$ and $J$ stand for the total
spin, relative orbital angular momentum and the total angular momentum of the initial $pp$ system,
respectively. The primed quantities refer to the final state. $l'$ denotes the orbital angular
momentum of the produced photon. We use the spectroscopic notation for the orbital angular
momenta.}
\begin{align}
E_{111} :&\ \ ^3P_1 \to{^1S_0s} \ , \ \ \ \ \ \ \ \ \ \ \ \ \ L_{000} :\ \ \ ^1S_0 \to{^1S_0p} \ , \nonumber \\
E_{111} :&\ \ ^3P_1 \to{^1S_0d} \ , \ \ \ \ \ \ \ \ \ \ \ \ \ E_{202} :\ \ ^1D_2 \to{^1S_0p} \ , \nonumber \\
M_{211} :&\ \ ^3P_2 \to{^1S_0d} \ , \ \ \ \ \ \ \ \ \ \ \ \ \ E_{202} :\ \ ^1D_2 \to{^1S_0f} \ , \nonumber \\
M_{213} :&\ \ ^3F_2 \to{^1S_0d} \ . \nonumber \\
%M_{213} :&\ \ ^3F_2 \to\ ^1S_0d \ , \ \ \ \ \ \ \ \ \ \ \ \ \ E_{404} :\ \ ^1G_4 \to\ ^1S_0f \ , \nonumber \\
%E_{313} :&\ \ ^3F_3 \to\ ^1S_0d \ . \nonumber \\
\label{PWS-spdf}
\end{align}
In the above list we have also displayed the corresponding multipole amplitudes to
which the specified partial waves contribute according to the classification given in the
previous section. The $^1S_0 \to{^1S_0p}$ partial wave, which enter the $L_{000}$
multipole, does not contribute in the present reaction because of the transversality
of the (real) photon. We therefore have four independent multipoles for $j\le2$.

Now, the spin matrix elements $M^{\perp/\parallel}_{SM_S}$ introduced in Sec.~\ref{sec:observables}
can be expressed in terms of the multipole amplitudes specified above. They are given
in Appendix B. Inserting those results (Eqs.~(\ref{eq:dd6},\ref{eq:dd7}) with the overall
normalization factor of $1/(4\pi)^{3/2}$ restored) into Eq.~(\ref{XSC-SCOR-unpol}),
we have for the unpolarized cross section
\begin{align}
\frac{d\sigma}{d\Omega} & =
 \frac{3}{(8\pi)^3}\Big\{|E_{111}+M2^+|^2 \Big. \nonumber \\
& \ \ \ \ \ \ \ \ + \left[2|E_{111}-M2^+|^2 - |E_{111}+M2^+|^2\right]\cos^2\theta_k \nonumber \\
& \ \ \ \ \ \ \ \ \Big. + \left[10|E_{202}|^2 + 4(|M2^-|^2 -|M2^+|^2) \right] \nonumber \\
&\ \ \ \ \ \ \ \ \ \ \ \ \ \ \ \ \ \ \ \ \ \ \ \ \ \ \ \ \ \ \ \ \ \ \ \ \ \ \ \
\times \cos^2\theta_k\sin^2\theta_k \Big\} \ , \nonumber \\
\label{eq:xsc-mult}
\end{align}
where $M2^{\pm}$ is defined as
\begin{equation}
M2^\pm \equiv M_{211}\pm \left(\frac{3}{2}\right)^{\mp \frac{1}{2}}M_{213} \ .
\label{eq:g14}
\end{equation}
Any deviation in the measured angular dependence from the above result is due to
higher $j$ states. In this connection, we note that the recent angular distribution data
from CELSIUS \cite{JW09} at  310 MeV proton incident energy reveals a
\begin{equation}
\frac{d\sigma}{d\Omega_k}  = \frac{3}{8\pi}\left[a + 3b\cos^2\theta_k
+ c\sin^2\theta_k\cos^2\theta_k\right]
\label{CELSIUS-data}
\end{equation}
dependence with $a = (2.3 \pm 0.3)$ nb, $b = (11.9 \pm 0.5)$ nb, and
$c < 2$ nb.  This indicates that, at this energy, the multipoles with $j > 2$
are relatively small and may be neglected.

If we now exclude the $^3F_2 \to{^1S_0d}$ transition ($M_{213}$ multipole),
as has been done in Ref.~\cite{WNA96}, Eq.~(\ref{eq:xsc-mult}) reduces to
\footnote{Recently, in Ref.~\cite{JW09}, an expression for the cross section
has been given which is at odds with the present result of Eq.~(\ref{eq:xsc2}).
They have corrected, however, in a later \textit{Erratum} \cite{JW09} which
agrees with the present result.}
\begin{align}
\frac{d\sigma}{d\Omega} &=
\frac{3}{(8\pi)^3}\Big\{|E_{111}+M_{211}|^2 \Big. \nonumber \\
&  + \left[3|E_{111}-M_{211}|^2 -2\left(|E_{111}|^2+|M_{211}|^2\right)\right]
\cos^2\theta_k \nonumber \\
& \Big. + 10|E_{202}|^2 \cos^2\theta_k\sin^2\theta_k \Big\} \ .
\label{eq:xsc2}
\end{align}
Given the smallness of the coefficient $c$ in Eq.~(\ref{CELSIUS-data}),
here, it is tempting to conclude that the $E_{202}$ multipole should be
very small at that energy. In fact, this is precisely the conclusion reached
in the analysis of Ref.~\cite{JW09} based on Eq.~(\ref{eq:xsc2}).
However, although this might be the case, one should be cautious in
drawing such a conclusion, as can be seen from a more general formula
given by Eq.~(\ref{eq:xsc-mult}).  The conclusion of the smallness of $E_{202}$
is warranted only if the $M_{213}$ multipole can be neglected.

In principle, the $E_{202}$ multipole can be determined directly if the
initial state spin-singlet cross section can be extracted:
\begin{equation}
\frac{d(^1\sigma)}{d\Omega}  = \frac{3}{(8\pi)^3}10|E_{202}|^2 \cos^2\theta_k\sin^2\theta_k \ ,
\label{S_xsc4aE}
\end{equation}
which means measuring not only the transverse but also the
longitudinal spin correlation coefficient $A_{zz}$ (cf. the first equality
in Eq.~(\ref{SPINMTX}). See, also Eq.~(D8) in Ref.~\cite{NL05}),
a quantity that is more challenging to be measured experimentally than the
transverse coefficients $A_{xx}$ and $A_{yy}$.

In terms of the multipoles with $j\le 2$, the analyzing power given by
Eq.~(\ref{Ay}) becomes
\begin{align}
\frac{d\sigma}{d\Omega} A_y & =
\frac{3}{4(4\pi)^3}\sqrt{10}Im[E_{202}(E_{111}-M2^+)^*]\nonumber \\
&\ \ \ \ \ \ \ \ \ \ \ \ \ \ \ \ \ \ \ \ \ \ \ \ \ \ \ \ \ \ \ \ \ \ \ \ \ \ \ \ \
\times \sin\theta_k\cos^2\theta_k \nonumber \\
& +\frac{3}{2(4\pi)^3}Im[M2^-(E_{111}+M2^+)^*]\sin\theta_k\cos\theta_k
\nonumber \\
& -\frac{3}{(4\pi)^3}Im[M2^-(M2^+)^*]\sin\theta_k\cos^3\theta_k \ .
\label{Ayr-mult}
\end{align}

Neglecting again the $^3F_2 \to{^1S_0d}$ partial wave state, the above
equation reduces to
\begin{align}
\frac{d\sigma}{d\Omega} A_y & =
\frac{3}{4(4\pi)^3}\sqrt{10}Im[E_{202}(E_{111}-M_{211})^*]\nonumber \\
&\ \ \ \ \ \ \ \ \ \ \ \ \ \ \ \ \ \ \ \ \ \ \ \ \ \ \ \ \ \ \ \ \ \ \ \ \ \ \ \ \
\times \sin\theta_k\cos^2\theta_k \nonumber \\
& +\frac{3}{2(4\pi)^3}Im[M_{211}(E_{111}+M_{211})^*]\sin\theta_k\cos\theta_k \ .
\label{Ayr1}
\end{align}
Comparing the results of Eqs.~(\ref{Ayr-mult},\ref{Ayr1}), it is clear that the angular
dependence  of the analyzing power (the $\sin\theta_k\cos^3\theta_k$ term)
can tell us about the size of the $^3F_2 \to{^1S_0d}$ matrix element, which has an
immediate consequence on the size of the $E_{202}$ multipole as has been discussed
above in connection with the (unpolarized) cross section.

As for the combinations of the (transverse) spin correlation coefficients in Eq.~(\ref{Atrans}),
we obtain
\begin{align}
\frac{d\sigma}{d\Omega} \left( A_{xx} + A_{yy} \right) & = \frac{3}{4(4\pi)^3} \left\{ 4|M2^-|^2 - 10|E_{202}|^2\right\}
\nonumber \\
&\ \ \ \ \ \ \ \ \ \ \ \ \ \ \ \ \ \ \ \ \ \ \ \ \ \ \times \sin^2\theta_k\cos^2\theta_k \ , \nonumber \\
\frac{d\sigma}{d\Omega} \left( A_{xx} - A_{yy} \right) & = \frac{3}{4(4\pi)^3} \left\{ 4|M2^+|^2\cos^2\theta_k \right.\nonumber \\
&\ \ \ \ \ \ \ \ \ \ \ \ \ \ \ \  \left. - |E_{111} + M2^+|^2 \right\} \sin^2\theta_k \ , \nonumber \\
\frac{d\sigma}{d\Omega} \left( 1 + A_{xx} \right) & = \frac{3}{4(4\pi)^3} \left\{ |E_{111}-M2^+|^2 \right. \nonumber \\
&\ \ \ \ \ \ \ \ \ \ \ \ \ \  \left. + 4|M2^-|^2 \sin^2\theta_k \right\} \cos^2\theta_k  , \nonumber \\
\frac{d\sigma}{d\Omega} \left( 1 + A_{yy} \right) & = \frac{3}{4(4\pi)^3} \left\{ |E_{111}+M2^+|^2\sin^2\theta_k
\right. \nonumber \\
& \ \ \ \ \ \ \ \ \ \ \ \ \left. + |E_{111}-M2^+|^2\cos^2\theta_k \right. \nonumber \\
& \ \ \ \ \ \ \ \ \ \ \ \  + \left. 4\left( |M2^-|^2 + |M2^+|^2 \right) \right. \nonumber \\
&\ \ \ \ \ \ \ \ \ \ \ \ \ \ \ \ \ \ \ \ \ \ \ \left. \times \sin^2\theta_k\cos^2\theta_k \right\} \ ,
\label{Atrans-2}
\end{align}
which reveal that the measurements of the transverse spin correlation coefficients
may be used to extract the multipole amplitudes involved. In particular, from the
angular distributions in the second and third equalities, one can extract $|M2^\pm|$
and the combinations $|E_{111} \pm M2^+|$. Then, the first equality can be used to
determine $|E_{202}|$.  Furthermore, we have $|E_{111}+ M2^+|^2 - |E_{111}-M2^+|^2
=4Re[E_{111}(M2^+)^*]$. The fourth equality may be used to check for consistency of
restricting to multipoles with $j\le 2$. Any deviation in the measured angular dependences
from the above results is an indication of the contribution from higher multipoles.

\section{Summary}
\label{sec:summary}

In the present work, we have derived the most general spin structure of the $NN$
bremsstrahlung amplitude where the two nucleons in the final state is restricted to
the $^1S_0$ state. The structure is consistent with general symmetry principles.
The coefficient multiplying each spin operator is expressed in terms of a linear
combination of the partial wave matrix elements. This is useful in analyses based
on partial wave expansion. These partial wave matrix elements have been related
to more familiar multipole amplitudes. This allows to study the hard bremsstrahlung
in $NN$ collisions in terms of either the partial wave or multipole amplitudes.

Based on the general spin structure obtained, we have shown that there are only four
independent spin matrix elements in the present reaction process which is a direct
consequence of reflection symmetry in the reaction plane. It requires at least eight
independent observables to determine uniquely these spin matrix elements. The
magnitude of them can be determined from the (diagonal) spin correlation coefficients
in conjunction with the unpolarized cross section (cf.~Eq.(\ref{SPINMTX})). The
analyzing power, together with the (off-diagonal) spin correlation coefficient, determines
the relative phase of a combination (cf.~Eqs.~(\ref{Axz},\ref{Ay})) of the spin matrix
elements involving two independent photon polarization states. This combination may
be disentangled if one measures both the analyzing power and off-diagonal spin
correlation coefficient together with the polarization of the photon. Obviously, such an
experiment will be extremely challenging by any standard.

We have also performed a partial-wave analysis of the
$p + p \to pp(^1S_0) + \gamma$ reaction considering the multipoles with the total
angular momentum $j\le 2$. It has been shown that these multipoles (or some
combination of them) can be determined from the measurements of the transverse
spin correlation coefficients, $A_{xx}$ and $A_{yy}$, and the analyzing power, $A_y$,
in conjunction with the unpolarized cross section. These observables can be measured
at the currently existing facilities such as COSY-ANKE at J\"ulich.

Finally, the results of the present study should be useful for the on going experimental
efforts \cite{ANKE1,ANKE2,JW09} to investigate the hard bremsstrahlung production
in $pp$ collisions as well as in providing some theoretical guidance to the future
experiments in this area.

%---------------------
\vspace{2em}
\noindent
{\bf Acknowledgment:}\\
%---------------------
\noindent
We thank Colin Wilkin and Ulf Mei\ss ner for a careful reading of the manuscript. This work
is supported by the FFE grant No. 41788390 (COSY-058).

\section{Appendix A}
In this appendix we will determine the coefficients $a_{lL}$, $b_{lL}$
in Eq.(\ref{AOPER}).

Taking the scalar product of the last equality in Eq.(\ref{AOPER}) with
$[\hat p \otimes \hat n_2]^2$ and $[\hat k \otimes \hat n_2]^2$,
respectively, we have
\begin{eqnarray}
2 r & = & a_{lL} + \cos\theta_k b_{lL} \nonumber \\
2 t & = & \cos\theta_k a_{lL} + b_{lL}  \ ,
\label{a6}
\end{eqnarray}
where
\begin{eqnarray}
r &\equiv & [Y_{L}(\hat p) \otimes Y_l(\hat k)]^2 \cdot [\hat p \otimes \hat n_2]^2 \nonumber \\
t &\equiv & [Y_{L}(\hat p) \otimes Y_l(\hat k)]^2 \cdot [\hat k \otimes \hat n_2]^2 \ .
\label{a7}
\end{eqnarray}
In order to arrive at Eq.(\ref{a6}), we have also made use of the results
\begin{eqnarray}
[\hat p \otimes \hat n_2]^2 \cdot [\hat p \otimes \hat n_2]^2  & = & \frac{1}{2} \nonumber \\
\small[\hat p \otimes \hat n_2]^2 \cdot [\hat k \otimes \hat n_2]^2  & = & \frac{1}{2} \cos\theta_k \ .
\label{a8}
\end{eqnarray}
Eq.(\ref{a6}) can be readily inverted to yield
\begin{eqnarray}
a_{lL}  & = & \frac{2}{\sin^2\theta_k} \left(r - t \cos\theta_k \right) \ , \nonumber \\
b_{lL}  & = & \frac{2}{\sin^2\theta_k} \left(t - r \cos\theta_k \right) \ .
\label{a9}
\end{eqnarray}

With the quantization axis $\hat z$ chosen along $\hat p$, the quantities $r$ and $t$
defined in Eq.(\ref{a7}) can be expressed as
\begin{align}
r & = - i \frac{1}{4\pi} \frac{[Ll]}{\sqrt{l(l+1)}} (L0l1|21) P^1_{l}(\hat k \cdot \hat p)
\nonumber \\
t & = i \frac{5\sqrt{2}}{4\pi} (-)^L \frac{[L]^2}{\sqrt{L(L+1)}} (10L0|l0)
(11L-1|L0) \nonumber \\
&\ \ \ \ \ \ \ \ \ \ \ \ \ \ \ \ \ \ \ \ \ \ \ \
\times   \begin{Bmatrix}
  1 & L & L \\   l & 2 & 1
  \end{Bmatrix}
P^1_{L}(\hat k\cdot \hat p) \ .
\label{a10}
\end{align}

If we consider the partial wave states with $j\le 2$, the
following $a_{lL}$ and $b_{lL}$ are required :
\begin{align}
a_{21} &= 0 \ , \nonumber \\
a_{23} &= -\frac{i}{4\pi}\frac{5}{2}\sqrt{15}\sin\theta_k\cos\theta_k \ , \nonumber \\
b_{21} &= \frac{i}{4\pi}\sqrt{15}\sin\theta_k \ , \nonumber \\
b_{23} &= \frac{i}{4\pi}\frac{\sqrt{15}}{2}\sin\theta_k \ .
\label{a11}
\end{align}
\section{Appendix B}

Here, we give the spin matrix elements in terms of the multipole amplitudes with $j\le 2$.
Using the short-hand notation
$M^{\lambda'}_{S\lambda} \equiv \braket{ 1\lambda'|\hat M(\vec k, \vec p\ '; \vec p)|S\lambda}$,
and leaving out the common factor of $1/(4\pi)^{3/2}$ for the moment, we have from
Eq.~(\ref{eq:g9}),
\begin{align}
&M^{\pm 1}_{00} = d^2_{0,\pm 1}(\theta_k)\sqrt{5}E_{202} =
\pm \sqrt{\frac{3}{2}}\sqrt{5}E_{202}\sin\theta_k\cos\theta_k \ , \nonumber \\
&M^{\pm 1}_{10} = d^1_{0,\pm 1}(\theta_k)(1010|10)\sqrt{3}E_{111} \nonumber \\[1ex]
& \ \ \ \ \ \ \
\pm d^2_{0,\pm 1}(\theta_k)\left[(1010|20)\sqrt{3}M_{211} + (3010|20)\sqrt{7}M_{213}\right]
\nonumber \\
& \ \ \ \ \ \ = \sqrt{\frac{3}{2}}\sqrt{2}\left[M_{211} - \sqrt{\frac{3}{2}}M_{213}\right]
\sin\theta_k\cos\theta_k \ , \nonumber \\
\label{eq:dd2}
\end{align}
\begin{align}
& M^{\pm 1}_{1,1} = d^1_{1,\pm 1}(\theta_k)(1011|11)\sqrt{3}E_{111} \nonumber \\
& \ \ \ \ \ \
\pm d^2_{1,\pm 1}(\theta_k)\left[(1011|21)\sqrt{3}M_{211} + (3011|21)\sqrt{7}M_{213}\right]
\nonumber \\
& \ \ \ \ \ \ \ = -\frac{1}{2}\sqrt{\frac{3}{2}} E_{111}(1\pm\cos\theta_k) \nonumber \\
& \ +\frac{1}{2}\sqrt{\frac{3}{2}}\left[M_{211} + \sqrt{\frac{2}{3}}M_{213}\right]
(2\cos^2\theta_k\pm\cos\theta_k-1) \ , \nonumber \\
\label{eq:dd3}
\end{align}
\begin{align}
& M^{\pm 1}_{1,-1} = d^1_{1,\pm 1}(\theta_k)(101-1|1-1)\sqrt{3}E_{111} \nonumber \\
&\ \ \ \ \ \ \ \ \ \
\pm d^2_{-1,\pm 1}(\theta_k)\left[(101-1|2-1)\sqrt{3}M_{211} \right. \nonumber \\
& \ \ \ \ \ \ \ \ \ \ \ \ \ \ \ \ \ \ \ \ \ \ \ \ \ \ \ \
\left. + (301-1|2-1)\sqrt{7}M_{213}\right]
\nonumber \\
& \ \ \ \ \ \ \ \ \ \ = \frac{1}{2}\sqrt{\frac{3}{2}} E_{111}\left(1\mp\cos\theta_k\right) \nonumber \\
& -\frac{1}{2}\sqrt{\frac{3}{2}}\left[M_{211} + \sqrt{\frac{2}{3}}M_{213}\right]
\left(2\cos^2\theta_k\mp\cos\theta_k-1\right) \ .
\label{eq:dd4}
\end{align}

The spin matrix elements $M^{\pm 1}_{S\lambda}$ given above are related
to the spin matrix elements $M^{\perp/\parallel}_{SM_S}$ introduced in
Sec.~\ref{sec:observables} by
\begin{align}
M^{\pm 1}_{S\lambda} & = \mp \frac{1}{\sqrt{2}} \left[ M^\parallel_{SM_S} \mp i M^\perp_{SM_S}\right] \ , \nonumber \\
M^{\parallel}_{SM_S} & =  \frac{1}{\sqrt{2}} \left[ M^{-1}_{S\lambda} - M^{+1}_{S\lambda}\right] \ , \nonumber \\
M^{\perp}_{SM_S} & =  \frac{-i}{\sqrt{2}} \left[ M^{-1}_{S\lambda} + M^{+1}_{S\lambda}\right] \ ,
\label{eq:gdd5}
\end{align}
with $\lambda = M_S$.

Using Eqs.~(\ref{eq:dd2},\ref{eq:dd3},\ref{eq:dd4}), we then have
\begin{align}
M^\parallel_{00} & = \sqrt{\frac{3}{2}}\sqrt{10}E_{202}\sin\theta_k\cos\theta_k \ , \nonumber \\
M^\perp_{00} & = 0\ ,
\label{eq:dd6}
\end{align}
for the spin singlet $\to$ singlet transitions, and
\begin{align}
M^\parallel_{10} & = 0 \ , \nonumber \\
M^\perp_{10} & =  -2\sqrt{\frac{3}{2}}\left[M_{211} - \sqrt{\frac{3}{2}}M_{213}\right]
\sin\theta_k\cos\theta_k \ , \nonumber \\
M^\parallel_{1\pm} & = \frac{1}{\sqrt{2}}\sqrt{\frac{3}{2}} \left( E_{111} -
\left[M_{211} + \sqrt{\frac{2}{3}}M_{213}\right] \right) \cos\theta_k \ , \nonumber \\
M^\perp_{1\pm} & = \pm \frac{i}{\sqrt{2}}\sqrt{\frac{3}{2}} \left( E_{111} -
\left[M_{211} + \sqrt{\frac{2}{3}}M_{213}\right]\cos\theta_k \right) ,
\label{eq:dd7}
\end{align}
for spin triplet $\to$ singlet transition matrix elements. Note that here we see an
explicit realization of the properties of the spin matrix elements
$M^{\perp/\parallel}_{SM_S}$ exhibited in Eqs.~(\ref{spin-singlet},\ref{spin-triplet})
in Sec.~\ref{sec:observables}.

%%%%%%%%%%%%%%%%%%%%%%%%%%%%%%%
%%%%%% References

%%%%%%%%%%%%%%%%%%%%%%%%%%%%%%%
%%%%%%%%%%%%%%%%%%%%%%%%%%%%%%%
%
%%%%%%%%%%%%%%%%%%%%%%%%%%%%%%%
\end{document}